\documentclass{article}
\usepackage{spconf,amsmath,graphicx}
\usepackage{amsmath,graphicx}
\usepackage{amsthm,amsmath,amssymb}
\usepackage{mathrsfs}
\usepackage{booktabs}
\usepackage{threeparttable}
\usepackage{multirow,multicol}
\usepackage{makecell}
\title{Improving hybrid CTC/Attention end-to-end speech recognition with pretrained acoustic and language models}
\name{$^*$Keqi Deng$^{1,2,3}$, $^*$Songjun Cao$^{1}$, Yike Zhang$^1$, Long Ma$^1$ \thanks{ $^*$ All the authors contributed equally to this work.}}
\address{
  $^1$Tencent Cloud Xiaowei, Beijing, China\\
  $^2$Institute of Acoustics, Chinese Academy of Sciences, China\\
  $^3$University of Chinese Academy of Sciences, China}
\begin{document}
%\ninept
%
\maketitle
\begin{abstract}
Recently, self-supervised pretraining has achieved impressive results in end-to-end (E2E) automatic speech recognition (ASR). However, the dominant sequence-to-sequence (S2S) E2E model is still hard to fully utilize the self-supervised pre-training methods because its decoder is conditioned on acoustic representation thus cannot be pretrained separately. In this paper, we propose a pretrained Transformer (Preformer) S2S ASR architecture based on hybrid CTC/attention E2E models to fully utilize the pretrained acoustic models (AMs) and language models (LMs). In our framework, the encoder is initialized with a pretrained AM (wav2vec2.0). The Preformer leverages CTC as an auxiliary task during training and inference. Furthermore, we design a one-cross decoder (OCD), which relaxes the dependence on acoustic representations so that it can be initialized with pretrained LM (DistilGPT2). Experiments are conducted on the AISHELL-1 corpus and achieve a $4.6\%$ character error rate (CER) on the test set. Compared with our vanilla hybrid CTC/attention Transformer baseline, our proposed CTC/attention-based Preformer yields $27\%$ relative CER reduction. To the best of our knowledge, this is the first work to utilize both pretrained AM and LM in a S2S ASR system.                                                       
\end{abstract}
\begin{keywords}
automatic speech recognition, end-to-end, sequence-to-sequence, pretraining
\end{keywords}
\section{Introduction}
\label{sec:intro}
End-to-end (E2E) automatic speech recognition (ASR) models integrate three parts of pipeline ASR methods: acoustics, pronunciation, and language into one \cite{graves2014towards,chan2016listen,watanabe2017hybrid} and directly transcribe input speeches into their corresponding transcripts. Thanks to large quantities of labeled training data, E2E ASR models outperform the pipeline methods on most public corpora \cite{8682586}. However, in many scenarios, labeled data is much harder to obtain than unlabeled data \cite{baevski2020wav2vec}, which is part of the reason why E2E ASR models still fall behind mature industry deployed ASR models. Pretraining can help alleviate this problem \cite{9398531} and self-supervised learning has gain success in numeral machine learning fields like computer vision \cite{DBLP:conf/iccv/DoerschGE15}, natural language processing (NLP) \cite{DevlinCLT19}, and ASR \cite{Schneider2019, keqi2021, songjun2021}.

However, although wav2vec2.0 \cite{baevski2020wav2vec} has been verified to work well, it has limited effect on attention-based S2S ASR models \cite{9398531}.
Furthermore, the decoder of S2S ASR models cannot be pre-trained separately because of its dependency on acoustics representations. And it is hard to utilize pre-trained LMs like BERT \cite{DevlinCLT19} or GPT2 \cite{radford2019language} for parameter initialization due to architecture mismatch \cite{9414080}. Therefore, how to efficiently utilize the pretrained acoustic models (AMs) and language models (LMs) in dominant S2S ASR models still remains a valuable challenge.

In this paper, we design a pretrained Transformer \cite{Vaswani2017} (Preformer) S2S ASR architecture based on the state-of-the-art (SOTA) hybrid connectionist temporal classification and attention (CTC/attention) E2E ASR model \cite{CTC-Attention-ACL-2017} to fully utilize those well-known pretrained AMs and LMs. In the CTC/attention-based Preformer, the encoder is initialized with a pretrained wav2vec2.0 \cite{baevski2020wav2vec}. The Preformer utilizes the CTC \cite{CTC_Graves_2006} branch to help the encoder converge better during training and to consider all possible time boundaries during beam searching.
In addition, we propose a one-cross decoder (OCD), which 
relaxes the dependence on acoustic representations. OCD is initialized by a pretrained DistilGPT2 \cite{wolf-etal-2020-transformers} and employed as the decoder of the Preformer.
Our experiments show that our proposed Preformer achieves 4.6\% character error rate (CER) on the test set of AISHELL-1 \cite{8384449} corpus. Compared with our vanilla hybrid CTC/attention Transformer baseline, the Preformer yields $27\%$ relative CER reduction\par

The rest of this paper is organized as follows. In Section~\ref{sec:Relate}, we introduce the related works.
In Section~\ref{sec:proposed}, we describe the CTC/attention-based Preformer. The experiments and conclusions are presented in Sections~\ref{sec:experiments} and \ref{sec:con}, respectively.\par

\begin{figure*}[th]
    \centering
    \includegraphics[width=160mm]{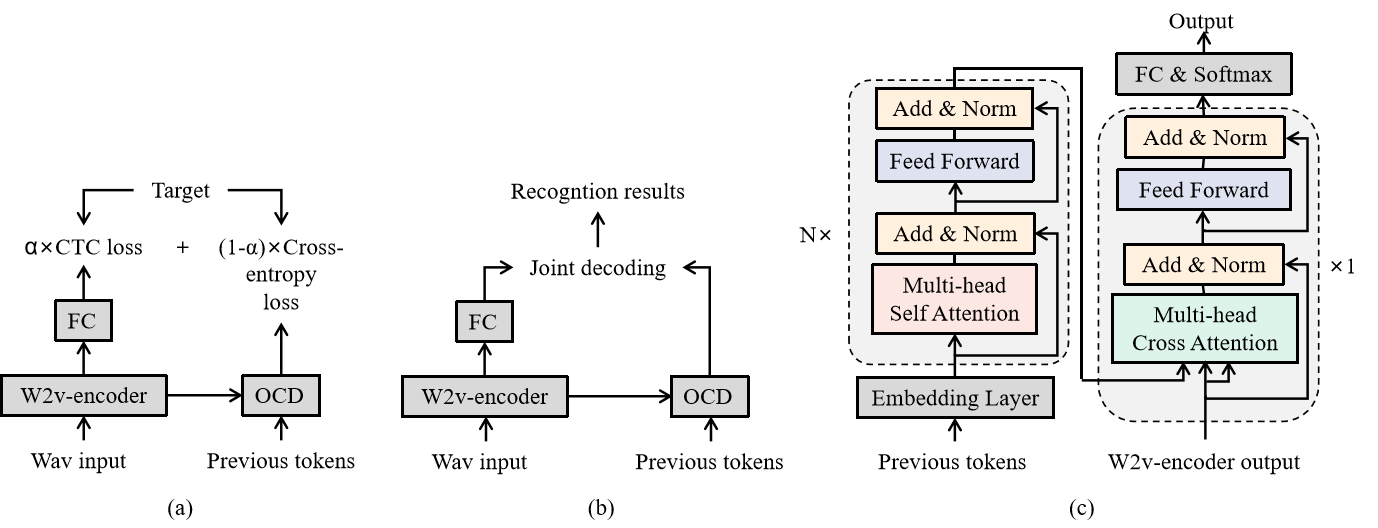}
    \caption{Illustration of the proposed CTC/attention-based Preformer architecture. (a) is the training process, (b) is the decoding process, and (c) is the structure of the proposed OCD.}
    \label{preformer}
\end{figure*}

\section{Related works}
\label{sec:Relate}
Self-supervised pretraining has gained success in E2E ASR tasks, recently. Wav2vec \cite{Schneider2019} learns the representations of raw speech through a self-supervised context-prediction task. In addition, Baevski et al. design vq-wec2vec \cite{DBLP:conf/iclr/BaevskiSA20} that learns discrete speech representations. Furthermore, Baevski et al. \cite{baevski2020wav2vec} further propose wav2vec 2.0, which learns the discrete speech units with contextual representations end-to-end. 

LM pretraining has become a paradigm for learning general language representations in NLP fields \cite{howard-ruder-2018-universal}. BERT is one such well-known pretrained LM \cite{DevlinCLT19}, which uses a Transformer encoder to build a text representation. Unlike BERT, GPT2 \cite{radford2019language} is composed of multi-layer unidirectional Transformer decoders. GPT2 only relies on past context to generate data representations and is well known as a language generator \cite{DBLP:journals/corr/abs-2004-02251}. And Sanh et al. propose a knowledge distillation method to compress the BERT into DistilBERT \cite{DBLP:journals/corr/abs-1910-01108}. This method is also used to compress GPT2 into DistilGPT2 \cite{wolf-etal-2020-transformers}.

LM plays an important part in ASR \cite{9414575}. Previous works like shallow fusion \cite{Kannan2018} and cold fusion \cite{Sriram2018} aim to combine an auto-regressive LM with a S2S ASR model, which is randomly initialized. To leverage the power of pretrained LM like BERT, many works concentrate on knowledge distillation of BERT for ASR \cite{futami2020distilling,9380383}, while the progress is limited \cite{DBLP:journals/corr/abs-2104-04805}. More recently, some studies aim to integrate the BERT into a non-autoregressive ASR models \cite{DBLP:journals/corr/abs-2104-04805,9398531}. However, how to efficiently utilize the pretrained LM like BERT in a dominant S2S ASR model is still an unsolvable challenge. 
\section{CTC/attention-based Preformer}
\label{sec:proposed}
The structure of the CTC/attention-based Preformer is shown in Fig.~\ref{preformer}, where w2v-encoder denotes the encoder that is initialized with a pretrained wav2vec2.0, and FC represents a fully connected layer.
The CTC/attention-based Preformer consists of a w2v-encoder, OCD, and an additional FC applied after the w2v-encoder for the CTC task. As shown in Fig.~\ref{preformer} (c), our proposed OCD has only one multi-head cross-attention mechanism and its left part is initialized with a pretrained DistilGPT2.
It should be noted that in the Preformer, only the FC and the right part of OCD in the Fig.~\ref{preformer} (c) do not receive any pretraining.

\subsection{Encoder (w2v-encoder)}
Many works try to use wav2vec2.0 to initialize the encoder of a vanilla Transformer based S2S ASR model, but the results obtained are very limited \cite{DBLP:journals/corr/abs-2012-12121, 9398531}. In this paper, we propose the CTC/attention-based Preformer and we believe that this hybrid architecture is a better choice to utilize the wav2vec2.0 in S2S ASR models.

The w2v-encoder of Preformer contains a CNN based feature encoder and a Transformer based context network. It converts raw speech input into acoustic representations. During fine-tuning, we only update the parameters of the Transformer based context network following the wav2vec2.0 \cite{baevski2020wav2vec}. 
Fine-tuning wav2vec2.0 on labeled data with CTC objective \cite{CTC_Graves_2006} has been well verified \cite{9414310, 9414227, 9414641}. However, according to the work \cite{9398531}, the results of fine-tuning wav2vec2.0 based on a vanilla Transformer S2S ASR model with cross-entropy criterion can only achieve a very limited result.
%is even worse than a CTC-based model. 
Therefore, in the Preformer, we apply an additional fully connected layer after the w2v-encoder and utilize the CTC objective as an auxiliary loss to help the w2v-encoder converge better during training. The training objective function $\mathcal{L}_{mtl}$ is:
\begin{equation}
    \mathcal{L}_{mtl}=\lambda \mathcal{L}_{{ctc}} + (1\!-\!\lambda)\mathcal{L}_{{ce}},
    \label{mot}
\end{equation}
where $\mathcal{L}_{ctc}$ and $\mathcal{L}_{ce}$ represent the CTC loss and cross-entropy loss, respectively. $\lambda$ is the weight of the CTC branch during joint training. Furthermore, during inference, we use the CTC branch to consider all possible time boundaries and improve the ability of the Preformer to recognize long utterances. And the decoding score is:
\begin{equation}
    S=\mu S_{{ctc}} + (1\!-\!\mu)S_{{ocd}},
\end{equation}
where $\mu$ represents the weight of CTC branch score $S_{{ctc}}$, and $S_{{att}}$ is the score predicted by OCD.

\subsection{One-cross decoder (OCD)}
Recently, how to use unpaired text data that is much easier to collect compared with paired speech-text data to improve S2S architecture has attracted more and more attention \cite{9053281}.
They try to solve the problem that the decoder depends on the encoder output and thus cannot be separately pretrained \cite{9414080, ramachandran-etal-2017-unsupervised}. For example, Gao et al. \cite{9414080} pretrain the decoder on unpaired text data with empty or artificial states instead of real encoder states, which not only fails to use those already pretrained LMs like GPT2, but the result is very limited due to the input state mismatch between pre-training and fine-tuning.

In the vanilla Transformer architecture, the decoder consists of a stack of $M$ identical layers \cite{Vaswani2017}. Each layer contains 3 sub-layers: a multi-head self-attention mechanism, a multi-head cross-attention mechanism, and a fully connected feed-forward module \cite{Vaswani2017}. 
%Suppose the encoder output as $E=({e}_1, \dots, {e}_T)$ and the decoder states as $D=({d}_1, \dots, {d}_L)$. The multi-head cross attention is calculated as follows:
%\begin{equation}
%    head_{i} = softmax\left( {\frac{(QW_{i}^{Q})(KW_{i}^{K})^{T}}{\sqrt{d_{m}}}} \right)(VW_{i}^{V}),
%\end{equation}
%\begin{equation}
%    output = concat\left( {head_{1},\ldots,head_{h}} \right)W^{O},
%\end{equation}
%where query matrix $Q$ equals $D$, matrices $K$ and $V$ equal $E$.
%And $W_{i}^{Q}\in\mathbf{R}^{d_{{model}} \times d_m}$, $W_{i}^{K}\in\mathbf{R}^{d_{{model}} \times d_m}$, and $W_{i}^{V}\in\mathbf{R}^{d_{{model}} \times d_m}$ are projection matrices for ${Q}$, ${K}$, and ${V}$ respectively, and $d_{{model}}$ is the model dimension.
The acoustic encoder output is fed into the multi-head cross-attention mechanism for $M$ times, which is why the decoder is so dependent on the acoustic encoder output and hard to alleviate. 

To relax the dependence on acoustic representations, we propose
%However, we believe that if there is a sufficiently high-quality text representation, performing cross-attention for only one time is enough. A high-quality text representation fully contains contextual information, which is beneficial for the accurate recognition of the ASR system. 
%Although some ambiguities caused by similar semantics may still be encountered in the next word prediction, as long as the text representation is combined with the high-quality acoustic encoder output through cross-attention mechanism once, the model can avoid the ambiguity and get correct predictions. 
%Therefore, we propose 
the OCD, which gets its name because it only performs a cross-attention mechanism once and is shown in Fig.~\ref{preformer} (c). 
we believe that performing cross-attention for only one time is enough with a sufficiently high-quality text representation, which fully contains contextual information and is beneficial for the accurate recognition of the ASR system. Although some ambiguities caused by similar semantics may still be encountered in the next word prediction, as long as the text representation is combined with the high-quality acoustic encoder output through cross-attention mechanism once, the model can avoid the ambiguity and get correct predictions. 

We denote the identical layer that contains a multi-head self-attention mechanism and feed-forward module as self layer, and that consists of a  multi-head cross-attention mechanism and feed-forward module as cross layer. The OCD consists of an embedding layer, a stack of $N$ self layers, and 1 cross layer. Since the OCD does not condition on the acoustic representation except for the last cross layer, we can use a pretrained LM like DistilGPT2 to initialize the parameters of the OCD before the cross layer. 

With a high-quality acoustic representation got from w2v-encoder and a context-aware text representation got from OCD, the Preformer can easily achieve a good recognition performance after fine-tuning.

\section{Experiments}
\label{sec:experiments}
\subsection{Corpus}
We evaluate the CTC/attention-based Preformer on the Mandarin AISHELL-1 corpus \cite{8384449}, which contains 178 hours of Mandarin speech. 
%The AISHELL-1 consists of about 120000 utterances training set and about 7000 utterances test set for testing. And its development set contains around 14000 utterances.
We also pretrain a Mandarin wav2vec2.0 base model using the training set of AISHELL-2 corpus \cite{DBLP:journals/corr/abs-1808-10583} without transcriptions.

\subsection{Model descriptions}
We use the ESPnet2 toolkit \cite{watanabe2018espnet} to build both the vanilla hybrid CTC/attention Transformer baseline and our proposed hybrid CTC/attention-based Preformer. For acoustic input,
we employ 80-dimensional filter banks for the baseline model and waveform speech data for the Preformer. As for the text output, we use 4230 Chinese characters with 3 non-verbal symbols as the modeling units. 
%\subsubsection{CTC/attention-based Transformer baseline}

We build the baseline model with a 12-layer encoder and a 6-layer decoder followed by ESPnet2 recipe \cite{watanabe2018espnet}. For fair comparison, we set the attention layer output dimension to 768 and the inner dimension of feed-forward layer to 3072 for both encoder and decoder. The attention heads for the encoder and decoder are 8 and 12, respectively.
%except that we increase the model dimension to 768. 
And an additional FC layer with 4233 dimensions is directly applied after the encoder as the CTC branch. 
%The total number of model parameters is 168.05M.

As for the CTC/attention-based Preformer, we first pretrain a Mandarin wav2vec2.0 base model (i.e., 7 convolution layers and 12 Transformer layers with 768 attention dimensions, 3072 feed-forward dimension, and 8 heads) on the training set of AISHELL-2 \cite{DBLP:journals/corr/abs-1808-10583} without transcriptions followed by Fairseq recipe \cite{ott2019fairseq}, in which the convolution layers are fixed during fine-tuning. And we use this Mandarin wav2vec2.0 model
to initialize the w2v-encoder. An additional FC is applied after the w2v-encoder as the CTC branch like the baseline model.
The OCD is composed of an embedding layer, 6 self layers, and 1 cross layer with 768 attention dimensions, 3072 feed-forward dimension, and 12 heads, in which the embedding layer and the self layers are initialized by a pretrained DistilGPT2, i.e., uer/gpt2-distil-chinese-cluecorpussmall (6 layers with 768 dimension) \cite{zhao2019uer} provided by Huggingface Transformer Library \cite{wolf-etal-2020-transformers}. 
During fine-tuning, for the first 10000 updates, only the FCs and the cross layer are optimized, after which the Transformer layers of w2v-encoder are also updated. The parameters that are initialized by the DistilGPT2 are fixed during fine-tuning. 
%And the total number of model parameters is 158.6M.
%since our preliminary experiments show that this will lead to a better results.
\begin{table}[h]
  \caption{Details of the ASR models' parameter number.}
  %\caption{Comparison among different embedding generation networks for the perplexity and the character error rate (CER) on HKUST.}
  \setlength\tabcolsep{4pt}
  \label{para}
  \centering
  \begin{tabular}{c c c c c}
    \toprule
    {ASR Model} &{Encoder}&{Decoder}&{CTC Branch}&{Total} \\
    \midrule
    baseline &101.6M&63.2M&3.3M&168.1M\\
    Preformer &94.4M&61.0M&3.3M&158.7M\\
    \bottomrule
  \end{tabular}
\end{table}

Details of both models' parameter number are shown in Table~\ref{para}.
We train both models by the Adam optimizer with warmup learning rate schedule (25000 warm steps) \cite{DBLP:journals/corr/GoyalDGNWKTJH17} for 20 epochs with 3 early stopping patience. We set the weight $\lambda$ of the CTC branch during joint training to 0.3. We also train an external Transformer LM followed by ESPnet2 recipe \cite{watanabe2018espnet}.
We average the models' parameters at the last 10 epochs to avoid overfitting. During joint decoding, we set the CTC-weight $\mu$ to 0.5 and the weight of external LM to 0.3, and the beam search size is 10.
\subsection{Experimental results}
We compare our proposed CTC/attention-based Preformer with our hybrid CTC/attention Transformer baseline model and other benchmark systems, and the results are shown in Table~\ref{tab:PC}. It should be noted that the official results of Transformer and Conformer provided by ESPnet1 and ESPnet2 \cite{watanabe2018espnet} are also based on the hybrid CTC/attention structure. The results show that our proposed Preformer greatly outperforms our Transformer baseline model and achieves a $27\%$ relative CER reduction on the test set, which proves the effectiveness of using the self-supervised pretrained AM and LM to improve the performance of S2S ASR model. Furthermore, compared with other benchmark systems, Preformer can even surpass the Conformer after using SpecAugment \cite{Park2019} data augmentation techniques, which is not employed in our experiments.
%But In our experiments, we do not utilize the SpecAugment technique, so this further shows the advantages of the Preformer.
\begin{table}[h]
  \caption{The character error rate (CER) (\%) of several benchmarks, our Transformer baseline, and our proposed Preformer on the AISHELL-1 corpus.}
  %\caption{Comparison among different embedding generation networks for the perplexity and the character error rate (CER) on HKUST.}
  \setlength\tabcolsep{5pt}
  \label{tab:PC}
  \centering
  \begin{tabular}{l c c}
    \toprule
    \multirow{2}{*}{ASR Model} &
    \multicolumn{2}{c}{CER} \\
    %\multirow{2}{*}{CER} \\
     & dev & test \\
    \midrule
    Kaldi (chain) \cite{povey2011kaldi} & -- &7.5\\
    Kaldi (nnet3) \cite{povey2011kaldi} & -- &8.7\\
    ESPnet2 (Transformer) \cite{watanabe2018espnet} & 6.1 &6.7\\
    ESPnet1 (Conformer) \cite{watanabe2018espnet} & 5.2 &5.8\\
    ESPnet2 (Transformer+SpecAug \cite{Park2019}) \cite{watanabe2018espnet} & 5.9 &6.4\\
    ESPnet2 (Conformer+SpecAug \cite{Park2019}) \cite{watanabe2018espnet} & 4.4 &4.7\\
    %\multicolumn{2}{c}{Baseline} & 50.76 & 24.9 &24.2\\
    \midrule
    Our Transformer baseline &5.9&6.3\\
    Our Preformer &\textbf{4.3}&\textbf{4.6}\\
    \bottomrule
  \end{tabular}
\end{table}
\subsection{Ablation studies on the OCD}
In the main experiments, we employ our proposed OCD as the decoder of the Preformer and fix the parameters that are initialized by the pretrained DistilGPT2 during fine-tuning. In this part, we conduct ablation studies to verify the effectiveness of the OCD and our fine-tuning recipe. The results are shown in Table~\ref{OCD}, where OCD-1 and OCD-3 represent setting the last 1 self layer and 3 self layers trainable during fine-tuning, OCD-all means that all the parameters of OCD can be updated during fine-tuning, 
and OCD-no-init represents a randomly initialized OCD and does not use the pretrained DistilGPT2 to initialize the parameters.
%, in which all the parameters can be optimized.
TCD denotes two-cross decoder that is the same as OCD except for two cross layers.

The results in Table~\ref{OCD} show that our proposed OCD can effectively utilize the pretrained LM DistilGPT2, thus greatly outperform the vanilla Transformer decoder. And compared with the vanilla Transformer decoder, the self layer of OCD removes the multi-head cross-attention mechanism, so the OCD actually has fewer parameters (i.e., 2.2M parameters less in this experiments) than the vanilla Transformer decoder, but it still gets better results. In addition, TCD has one more cross layer than OCD but it does not improve the performance, which proves our previous point: performing cross-attention for only one time is enough with a sufficiently high-quality text representation.
Furthermore, we can see that fixing all the parameters that are initialized by the DistilGPT2 does perform best, especially in the Preformer where the w2v-encoder is employed. This is because we mainly utilize the DistilGPT2 to extract high-quality text representations, which can 
also improve the fine-tuning of w2v-encoder through the cross-modal learning of the cross layer.
%we only need the high-quality text representation extracted by the DistilGPT2 and do not need it to predict next words.
%A fixed high-quality text representation can better help the fine-tuning of w2v-encoder through the cross-modal learning of the cross layer.
And fine-tuning the parameters that are initialized by the DistilGPT2 on our very limited labeled dataset may damage the quality of the extracted text representation because of overfitting, thus reducing the final recognition performance.  
Finally, it can be seen from the last row of Table~\ref{OCD} that using the pretrained DistilGPT2 to initialize the OCD parameters is indeed beneficial to improve the accuracy of the S2S E2E ASR system.
\begin{table}[h]
  \caption{The CER (\%) of our ASR models with different decoders on the AISHELL-1 corpus.}
  %\caption{Comparison among different embedding generation networks for the perplexity and the character error rate (CER) on HKUST.}
  \setlength\tabcolsep{5pt}
  \label{OCD}
  \centering
  \begin{tabular}{c c c c c}
    \toprule
    \multirow{2}{*}{ASR Model} & \multirow{2}{*}{Encoder} & \multirow{2}{*}{Decoder} &
    \multicolumn{2}{c}{CER} \\
    %\multirow{2}{*}{CER} \\
    & & & dev & test \\
    \midrule
    baseline& Transformer & Transformer&5.9&6.3\\
    %\multicolumn{2}{c}{Baseline} & 50.76 & 24.9 &24.2\\
    \midrule
    proposed &Transformer&OCD&5.1&5.6\\
    proposed &Transformer&OCD-1&5.2&5.7\\
    proposed &Transformer&OCD-3&5.2&5.7\\
    proposed &Transformer&OCD-all&5.3&5.7\\
    \midrule
    %proposed &w2v-encoder&OCD-no-init&4.7&5.1\\
    %\midrule
    Preformer &w2v-encoder&OCD&\textbf{4.3}&\textbf{4.6}\\
    Preformer &w2v-encoder&OCD-1&4.5&4.9\\
    Preformer &w2v-encoder&OCD-3&4.7&5.0\\
    Preformer &w2v-encoder&OCD-all&4.7&5.1\\
    \midrule
    proposed &w2v-encoder&TCD&4.3&4.6\\
    proposed &w2v-encoder&OCD-no-init&4.9&5.2\\
    \bottomrule
  \end{tabular}
\end{table}

\subsection{Ablation studies on the w2v-encoder}
We also conduct ablation studies to verify the effectiveness of the w2v-encoder and the results are shown in Table~\ref{w2v}, where the w2v-encoder-En means using the English wav2vec2.0 base model provided by the Fairseq \cite{ott2019fairseq} to initialize the encoder. And the results show that the wav2vec2.0 is not only effective in the CTC-based ASR models, but also can help improve the recognition performance of the CTC/attention-based S2S ASR system. In addition, it can be seen that even directly using the English wav2vec2.0 model to initialize the encoder can still greatly outperform the vanilla Transformer encoder on the Mandarin ASR tasks, which also indicates that using cross-language pre-training method can help solve the ASR tasks in low-resource languages.
And after the Mandarin wav2vec2.0 model is used, the recognition performance is further improved, which is intuitive.
\begin{table}[h]
  \caption{The CER (\%) of our ASR models with different encoders on the AISHELL-1 corpus.}
  %\caption{Comparison among different embedding generation networks for the perplexity and the character error rate (CER) on HKUST.}
  \setlength\tabcolsep{5pt}
  \label{w2v}
  \centering
  \begin{tabular}{c c c c c}
    \toprule
    \multirow{2}{*}{ASR Model} & \multirow{2}{*}{Encoder} & \multirow{2}{*}{Decoder} &
    \multicolumn{2}{c}{CER} \\
    %\multirow{2}{*}{CER} \\
    & & & dev & test \\
    \midrule
    baseline& Transformer & Transformer&5.9&6.3\\
    %\multicolumn{2}{c}{Baseline} & 50.76 & 24.9 &24.2\\
    \midrule
    proposed &w2v-encoder&Transformer&4.7&5.0\\
    proposed &w2v-encoder-En&Transformer&5.1&5.3\\
    \midrule
    Preformer &w2v-encoder&OCD&\textbf{4.3}&\textbf{4.6}\\
    Preformer &w2v-encoder-En&OCD&4.6&4.8\\
    \bottomrule
  \end{tabular}
\end{table}

\subsection{Ablation studies on the use of DistilGPT2}
There are many ways to utilize pretrained LMs. The simplest and effective method is to fine-tune a pretrained LM as the external LM of an ASR system through shallow fusion \cite{Kannan2018}. In the main experiments, we train a 16-layer external vanilla Transformer LM followed by ESPnet2 recipe \cite{watanabe2018espnet}.
In this part, we additionally fine-tune a DistilGPT2 as the external LM to improve the baseline model and compare it with our Preformer. The results are shown in Table~\ref{lm}, from which we can see that fine-tuning a pretrained LM as the external LM through shallow fusion \cite{Kannan2018} can indeed improve the ASR performance, but this is limited. The OCD structure of the Preformer we proposed can more effectively utilize the power of the pre-trained LM in extracting high-quality text representation for the ASR system, thus improving the final ASR accuracy. Furthermore, using the DistilGPT2 as the external LM of the Preformer can further improve the recognition performance.
\begin{table}[h]
  \caption{The CER (\%) of our ASR models with different ways of using the DistilGPT2 on the AISHELL-1 corpus.}
  %\caption{Comparison among different embedding generation networks for the perplexity and the character error rate (CER) on HKUST.}
  \setlength\tabcolsep{2.3pt}
  \label{lm}
  \centering
  \begin{tabular}{c c c c c c}
    \toprule
    \multirow{2}{*}{ASR} & \multirow{2}{*}{Encoder} & \multirow{2}{*}{Decoder} & \multirow{2}{*}{LM}&
    \multicolumn{2}{c}{CER} \\
    %\multirow{2}{*}{CER} \\
    & & & & dev & test \\
    \midrule
    baseline& Transformer & Transformer&vanilla&5.9&6.3\\
    baseline& Transformer & Transformer&DistilGPT2&5.3&5.7\\
    %\multicolumn{2}{c}{Baseline} & 50.76 & 24.9 &24.2\\
    \midrule
    proposed &Transformer&OCD&vanilla&5.1&5.6\\
    proposed &Transformer&OCD&DistilGPT2&4.8&5.2\\
    \midrule
    Preformer &w2v-encoder&OCD&vanilla&4.3&4.6\\
    Preformer &w2v-encoder&OCD&DistilGPT2&\textbf{3.9}&\textbf{4.2}\\
    \bottomrule
  \end{tabular}
\end{table}

\section{Conclusion}
\label{sec:con}
In this paper, we explore utilizing the self-supervised pretrained AMs and LMs to improve the S2S E2E ASR models. And we propose a CTC/attention-based Preformer, in which the encoder is initialized with a pretrained wav2vec2.0 model. The Preformer utilizes the CTC branch to help the encoder converge better during training and to consider all possible time boundaries during beam searching.
We also design the OCD to relax the dependence on the encoder output so that most of it can be separately pretrained. The OCD is employed as the decoder of the Preformer and initialized with a pretrained DistilGPT2. Experiments results on the AISHELL-1 corpus show that the CTC/attention-based Preformer achieves a 4.6\% CER on the test set and yields $27\%$ relative CER reduction compared with our vanilla hybrid CTC/attention Transformer baseline. 

%In the future, we plan to build a streaming Preformer based on a streaming wav2vec2.0, and try to make a non-streaming Preformer based on conformer-based wav2vec2.0 to further improve the ASR accuracy.
\bibliographystyle{IEEEbib}
\bibliography{strings,refs}

\end{document}